\documentclass{ifacconf}

\usepackage{graphicx}      
\usepackage{natbib}        
\usepackage{amsmath,amsfonts,amssymb}
\usepackage{stmaryrd}
\usepackage{booktabs}
\usepackage[ruled]{algorithm}
\usepackage[algo2e]{algorithm2e}
\newlength\hrulethickness
\makeatletter
\renewcommand\fs@ruled{\def\@fs@cfont{\bfseries}\let\@fs@capt\floatc@ruled
  \def\@fs@pre{\hrule height 1.5\hrulethickness depth0pt \kern2pt}%
  \def\@fs@post{\kern2pt\hrule height 1.5\hrulethickness depth0pt \relax}%
  \def\@fs@mid{\kern2pt\hrule height 0.4\hrulethickness depth0pt \kern2pt}%
  \let\@fs@iftopcapt\iftrue}
\makeatother
\begin{document}
\begin{frontmatter}

\title{Weak Interactions Based System Partitioning Using Integer Linear Programming} 


\author[First]{Romain~Guicherd}
\author[First]{Paul~A.~Trodden} 
\author[First]{Andrew~R.~Mills}
\author[First]{Visakan~Kadirkamanathan}

\address[First]{Rolls-Royce~University~Technology~Centre, Department~of~Automatic~Control \& Systems~Engineering, University~of~Sheffield, Mappin~Street, Sheffield, S1~3JD, UK (e-mail:rguicherd1@sheffield.ac.uk; p.trodden@sheffield.ac.uk; a.r.mills@sheffield.ac.uk; visakan@sheffield.ac.uk).}

\begin{abstract}                
The partitioning of a system model will condition the structure of the controller as well as its design. In order to partition a system model, one has to know what states and inputs to group together to define subsystem models. For a given partitioning, the total magnitude of the interactions between subsystem models is evaluated. Therefore, the partitioning problem seeking for weak interactions can be posed as a minimization problem. Initially, the problem is formulated as a non-linear integer minimization that is then relaxed into a linear integer programming problem. It is shown within this paper that cuts can be applied to the initial search space in order to find the least interacting partitioning; only composed of controllable subsystems. Two examples are given to demonstrate the methodology.
\end{abstract}


\end{frontmatter}

\section{Introduction}
Systems models are widely used in control design especially with the development of techniques such as model predictive control (\cite{Rawlings2009}) (\cite{Maciejowski2002}). Systems are growing in size and complexity and they are in most cases composed of interacting subsystems (\cite{Scattolini2009}). For these large scale systems, the design of a centralized controller can be prohibitive due to the heavy computational resources required (\cite{Mayne2014}). Also, if the system is geographically spread out, communication delays between the centralized controller and the actuators and sensors arise. One way to solve this problem is to see the system as a concatenation of subsystems and to design local controllers for each subsystem. In a top-down approach the full model of the multivariable system is partitioned into subsystem models so that the decentralized controller can be designed. Decentralized control has been studied for decades and design procedures have been established (\cite{Siljak1991}) (\cite{Bakule2008}). However, the system model partitioning problem has been overlooked, often because the system is already composed of physical subsystems. Every subsystem model is defined by a set of states and  inputs. The weak interaction partitioning problem consist of defining these sets in order to minimize the coupling between subsystem models. For instance, strongly coupled subsystem models can emerge from the main system model, particularly within chemical plants (\cite{Stewart2010}) or heating systems (\cite{Morosan2010}). The ideal partitioning of a system model would yield completely decoupled subsystem models.

Defining the subsystems of a plant has been done in different ways in the past. One of the first methods employed to couple inputs and outputs was the relative gain array (\cite{Bristol1966}). This method is used to find the best pairing  at steady state between inputs and outputs and hence to choose the most relevant input to control a given output in a multi-input multi-output system. It can be seen as a response to the industrial need to control a multi-variable process as a combination of single variable processes. The relative gain array has been extended to the block relative gain, allowing for suitable pairing for block decentralized control (\cite{Manousiouthakis1986}) (\cite{Kariwala2003}). The extension of the relative gain array allows the design of multi-variable controllers in a decentralized way. However it only links inputs and outputs together and does not provide  a partitioning of the plant model. A technique similar to the relative gain array, is the Nyquist array method, allowing the design of single-input single-output controllers after rendering the model diagonaly dominant (\cite{Leininger1979}) (\cite{Chen2003}). System partitioning can be performed by seeking the least interacting groups. Another technique used for system decomposition and integration is the design structure matrix also known as the dependency structure matrix or interaction matrix (\cite{Browning2001}). This technique indicates the link between the elements it represents, moreover the links are directed. Elements along a row indicate that a contribution is provided to other elements whereas elements along a column indicates a dependency from other parts of the system. The attribution of weights within the interaction matrix is used in order to perform clustering and achieve system decomposition. Other work on decentralized control combined the controller design along with the controller topology, these two aspects are combined in the optimization function yielding a trade-off between the need for feedback links and the loss of performance compared to the centralized controller (\cite{Schuler2014}). Finally, other works have studied the actuator partitioning problem (\cite{Jamoom1998}) (\cite{Motee2003}). To the best of the authors' knowledge the problem addressing state space model partitioning has not been studied. Therefore this partitioning approach is a standalone work, making any comparison difficult.

In this paper, we propose an integer programming based approached to the problem of partitioning a system model into a set of non-overlapping but coupled subsystem models. The objective is to reduce the magnitude of the interactions between the subsystem models. Finally, cuts are added to rule out non-controllable partitionings in order for the algorithm to yield only controllable subsystem models.

The paper is organised as follows. Section 2 states the problem and section 3 introduces the required notations. Section 4 demonstrates how the problem can be relaxed into a linear integer programming problem. In section 5 the controllability cut principle is presented allowing to obtain controllable subsystems, section 6 explains the linear partitioning algorithm. In order to illustrate the algorithm section 7 includes some examples, finally section 8 concludes the paper.

\emph{Notation:} For $(a,b) \in \mathbb{N}^{2}$ such that $a<b$, the set $\llbracket a;b \rrbracket$ defines the set containing the integers from $a$ to $b$ included. The operator $\vert . \vert$ is used to denote the magnitude of a complex number when applied to a complex number. When applied to a matrix the magnitude operator is applied to all the matrix elements then summed. The operator $\lVert . \rVert_{2}$ defines the Euclidean norm for complexes, vectors and matrices. For a set $\mathbb{N}$, the notation $\mathbb{N}^{*}$ defines $\mathbb{N} / \{0\}$. The superscript $^{\intercal}$ represents the transpose of a vector or matrix. A matrix $B \in \mathbb{R}^{n \times m}$ will be noted $(b_{ij})_{(i,j) \in \llbracket 1;n \rrbracket \times \llbracket 1;m \rrbracket}$ and $(B_{kl})_{(k,l) \in \llbracket 1;N \rrbracket \times \llbracket 1;M \rrbracket}$, respectively for the element and block notations, with $N$ row blocks and $M$ column blocks. For all $n \in \mathbb{N}^{*}$ and $A \in \mathbb{R}^{n \times n}$, $trace(A)$ denotes the sum of all the diagonal elements of $A$. For all $(n,m) \in (\mathbb{N}^{*})^{2}$ and $A \in \mathbb{R}^{n \times m}$, $rank(A)$ denotes the dimension of the vector space spanned by the columns of $A$. For any $(i,j,k) \in (\mathbb{N}^{*})^{3}$ and $(A,B) \in \mathbb{R}^{i \times j} \times \mathbb{R}^{i \times k}$ the notation $[A | B]$ defines the matrix obtained by concatenating $A$ and $B$ horizontally. For any couple of integers $(i,j)\in \mathbb{N}^{2}$, $\delta_{ij}$ denotes the Kronecker delta function.

\section{Problem statement}

Given a linear time invariant controllable state space model defined by

\begin{equation} \label{eqn1}
\dot{x} = Ax + Bu
\end{equation}

where the matrix $A$ is the state matrix and the matrix $B$ is the input matrix respectively with the appropriate sizes for $N$ states and $M$ inputs, therefore, $x \in \mathbb{R}^{N}$ and $u \in \mathbb{R}^{M}$. Partitioning the system model (\ref{eqn1}) consists of decomposing the inputs as well as the states into groups representing subsystems. For a given number of partitions $P \in \llbracket 2;\min(N,M) \rrbracket$ and for any subsystem $p \in \llbracket 1;P \rrbracket$ the model $p$ can be expressed as follows

\begin{equation} \label{eqn2}
\dot{x}_{p} = A_{pp} x_{p} + B_{pp} u_{p} + \sum_{\substack{j=1 \\ j \neq p}}^{P} \big\{ A_{pj} x_{j} + B_{pj} u_{j}  \big\}
\end{equation}

with for all $p \in \llbracket 1;P \rrbracket$, $x_{p} \in \mathbb{R}^{N_{p}}$ and $u_{p} \in \mathbb{R}^{M_{p}}$ such that

\begin{subequations} \label{eqn3}
\begin{align} \label{eqn3a}
\sum_{p=1}^{P} N_{p} = N \\ \label{eqn3b}
\sum_{p=1}^{P} M_{p} = M
\end{align}
\end{subequations}

The weak interaction partitioning problem consists of minimizing the magnitude of the right-hand side sum in (\ref{eqn2}) for the subsystems while keeping each of them controllable. A non-overlapping condition for the states and the inputs is imposed by (\ref{eqn3}). The next section presents the decision variables, the constraints as well as the interaction metric necessary to formulate the weak interactions optimization problem.

\section{Weak interactions problem formulation}

\subsection{Decision variables}
A decision variable is associated to the couples formed by a group $p$ and a state $i$ as well as a group $p$ and an input $k$. All the decision variables are binary variables. They are organised in two grouping matrices, the state grouping matrix $\alpha \in \llbracket 0;1 \rrbracket^{P \times N}$ and the input grouping matrix $\beta \in \llbracket 0;1 \rrbracket^{P \times M}$. Therefore, the rows of $\alpha$ and $\beta$ represent the $P$ groups and the columns respectively represent the $N$ states and the $M$ inputs. For example with $P=3$ and $N=5$ a non-overlapping state grouping matrix could be

\begin{equation*}
\alpha = \left(
\begin{array}{ccccc}
 0 & 0 & 0 & 1 & 0 \\
 1 & 1 & 1 & 0 & 0 \\
 0 & 0 & 0 & 0 & 1
\end{array}
\right)
\end{equation*}

In this example, the first three states belong to the second group, the fourth state composes the first group and the last state is in the third group. Hence, a specific partitioning is represented by a pair of state and input non-overlapping grouping matrices. The columns of the grouping matrices are composed of zeros and a single one. The one is positioned in the row representing the group where the state or input belongs respectively for a state and an input non-overlapping grouping matrix. The next subsection presents the linear constraints restricting the decision variables $\alpha$ and $\beta$ in the integer optimization problem.

\subsection{Partitioning constraints}
The formulation of constraints on the decision variables is necessary in order for the algorithm to return a solution complying with the rules defining non-overlapping subsystem models.

\begin{itemize}
\begin{subequations} \label{eqn5}
\item[1.]Each state group contains at least a state, hence, no state group can be empty and the partitioning has the correct number of state groups
\begin{align}
\displaystyle \forall p \in \llbracket 1; P \rrbracket, \, \sum_{i=1}^{N} \alpha_{pi} \geq 1 \label{eqn5a}
\end{align}
\item[2.]Each input group contains at least an input, hence, no input group can be empty and the partitioning has the correct number of input groups
\begin{align}
\displaystyle \forall p \in \llbracket 1; P \rrbracket, \, \sum_{i=1}^{M} \beta_{pi} \geq 1 \label{eqn5b}
\end{align}
\item[3.]A state can be in only one state group, therefore, the multiple use of a state is prevented and the non-overlapping requirement is respected
\begin{align}
\displaystyle \forall i \in \llbracket 1; N \rrbracket, \, \sum_{p=1}^{P} \alpha_{pi} \leq 1 \label{eqn5c}
\end{align}
\item[4.]An input can be in only one input group, therefore, the multiple use of an input is prevented and the non-overlapping requirement is respected
\begin{align}
\displaystyle \forall i \in \llbracket 1; M \rrbracket, \, \sum_{p=1}^{P} \beta_{pi} \leq 1 \label{eqn5d}
\end{align}
\item[5.]Each state must belong to a state group, consequently, no state is left out of the optimization problem
\begin{align}
\displaystyle \forall i \in \llbracket 1; N \rrbracket, \, \sum_{p=1}^{P} \alpha_{pi} \geq 1 \label{eqn5e}
\end{align}
\item[6.]Each input must belong to a input group, consequently, no input is left out of the optimization problem
\begin{align}
\displaystyle \forall i \in \llbracket 1; M \rrbracket, \, \sum_{p=1}^{P} \beta_{pi} \geq 1 \label{eqn5f}
\end{align}
\end{subequations}
\end{itemize}

The constraints are expressed for the two grouping matrices, however only three different sets of constraints concern each type of grouping matrix. Because $\alpha$ and $\beta$ are arrays of binary variables a natural implicit constraint links $N,M$ and $P$.

\begin{equation} \label{eqn6}
1 < P \leq \min(N,M)
\end{equation}

Subsystem interactions can come from the state matrices or the input matrices, the next subsection presents how these interactions can be formulated firstly using the block matrix form and secondly using the state space model elements.

\subsection{Objective: minimizing subsystem interactions}
The first part of the interactions comes from the state matrices. The subsystem model (\ref{eqn2}) presents the couplings with the other subsystems in the form of a sum, this sum can be split into the state interactions and the input interactions. For a given number of partitions $P \in \llbracket 2;\min(N,M) \rrbracket$ and for any subsystem $p \in \llbracket 1;P \rrbracket$ the state interactions can be expressed by

\begin{equation} \label{eqn7}
J^{state}_{p} = \sum_{\substack{j=1 \\ j \neq p}}^{P} \lvert A_{pj} \rvert
\end{equation}

The expression written in block matrix form can also be represented using the state matrix elements as well as the state grouping matrix elements as follows

\begin{equation} \label{eqn8}
J^{state}_{p} = \sum_{i=1}^{N} \sum_{j=1}^{N} \alpha_{pi} \lvert a_{ij} \rvert (1-\alpha_{pj})
\end{equation}

The elements from the state grouping matrix are used here as boolean tests to take into account only the interactions acting on the subsystem $p$ and coming from the other subsystem states. A similar reasoning is applied to quantify the interactions coming from the input matrices

\begin{equation} \label{eqn9}
J^{input}_{p} = \sum_{\substack{j=1 \\ j \neq p}}^{P} \lvert B_{pj} \rvert
\end{equation}

In a similar fashion, (\ref{eqn9}) can also be represented using the input matrix elements as well as the elements of the state grouping matrix combined with the elements of the input grouping matrix, such that

\begin{equation} \label{eqn10}
J^{input}_{p} = \sum_{i=1}^{N} \sum_{k=1}^{M} \alpha_{pi} \lvert b_{ik} \rvert (1-\beta_{pk})
\end{equation}

Likewise, the elements from the state grouping matrix combined with the elements of the input grouping matrix are used as boolean tests to take into account only the interactions acting on the subsystem $p$ and coming from the other subsystem inputs. After having defined the two types of interactions, the full interaction metric can be calculated. Consequently, the last step is to pose the weak interactions optimization problem like it is presented within the next subsection.

\subsection{Weak interactions optimization problem}
The overall formulation of the weak interactions optimization problem is obtained by summing the interactions coming from the states and the inputs over the $P$ subsystems

\begin{equation} \label{eqn11}
J^{interaction} = \displaystyle \sum_{p=1}^{P} \big\{ J^{state}_{p} + J^{input}_{p} \big\}
\end{equation}

Hence, the integer optimization problem can be formulated using the same notation employed in (\ref{eqn8}) and (\ref{eqn10}) and is expressed as follows

\begin{equation} \label{eqn12}
\begin{aligned}
&\underset{\alpha ,\beta}{\text{min}} & & \displaystyle \sum_{p=1}^{P} \sum_{i=1}^{N} \bigg\{ \alpha_{pi} \Big[ \sum_{j=1}^{N} \big\{ \lvert a_{ij} \rvert (1-\alpha_{pj}) \big\} \\
& & & + \sum_{k=1}^{M} \big\{ \lvert b_{ik} \rvert (1-\beta_{pk}) \big\} \Big] \bigg\} \\
&\text{s. t.}
& & (\ref{eqn5})
\end{aligned}
\end{equation}

As it was demonstrated previously within this section the partitioning problem can be expressed as an integer optimization problem. However the cost function representing the interaction metric is non-linear, therefore the problem can be intractable. As it is presented within the next section a linear relaxation of the optimization problem (\ref{eqn12}) is made possible throughout the use of auxiliary variables.

\section{Linear relaxation of the weak interactions problem}

The weak interactions optimization problem formulated previously (\ref{eqn12}) can be turned into a linear optimization problem, this is made possible due to the introduction of auxiliary variables. Replacing a product of binary variables by an auxiliary variable is a well known technique that requires the use of new linear constraints (\cite{Williams2013}) (\cite{Bemporad1999}) (\cite{Cavalier1990}). Two auxiliary binary variables are created along with their linear constraints. $\gamma$ is the binary variable used to take into account the interactions coming from the state matrix. As it is presented in Table I and in (\ref{eqn13}), $\gamma$ is linked to $\alpha$ throughout four constraints. Indeed, four inequalities are necessary because of the four possible outcomes for the binary product $\alpha_{pi}(1-\alpha_{pj})$ in (\ref{eqn12}).

From top to bottom within Table I, the four different cases are, first when no states belong to the group $p$ then no interaction has to be accounted for. If the state $i$ is not in the group $p$ but the state $j$ is, then no interaction is accounted for as this will be taken into account in the symmetrical case. If the state $i$ belongs to the group $p$ and the state $j$ does not then an interaction is accounted for. The last case possible is when the two states $i$ and $j$ both belong to the group $p$, in this last scenario no interaction subsists as they are both in the same group.

\begin{table}[ht]
\centering
\normalsize
\caption{Auxiliary variable $\gamma$}
\begin{tabular}{ccccc}
\toprule[1.5pt]
\multicolumn{3}{c}{\textbf{Primary}} & {\textbf{Auxiliary}} \\
$\alpha_{pi}$ & $\alpha_{pj}$ & $\alpha_{pi}(1-\alpha_{pj})$ & $\gamma_{pij}$ \\
\cmidrule[0.4pt](lr){1-3} \cmidrule[0.4pt](lr){4-4}
0 & 0 & 0 & 0 \\
0 & 1 & 0 & 0 \\
1 & 0 & 1 & 1 \\
1 & 1 & 0 & 0 \\
\bottomrule[1.5pt]
\end{tabular}
\\[5pt]
\end{table}

The linear constraints for the auxiliary variable $\gamma$ are the following

\begin{subequations} \label{eqn13}
$\forall (p;i;j) \in \llbracket 1; P \rrbracket \times \llbracket 1; N \rrbracket \times \llbracket 1; N \rrbracket,$
\begin{align} \label{eqn13a}
\gamma_{pij} \leq \alpha_{pi} + \alpha_{pj} \\ \label{eqn13b}
\gamma_{pij} \leq 1 + \alpha_{pi} - \alpha_{pj} \\ \label{eqn13c}
\gamma_{pij} \geq \alpha_{pi} - \alpha_{pj} \\ \label{eqn13d}
\gamma_{pij} \leq 2 - \alpha_{pi} - \alpha_{pj}
\end{align}
\end{subequations}

In a similar way, $\delta$ is the auxiliary binary variable taking into account the interactions coming form the input matrix. This time the constraints are formed using $\alpha$ as well as $\beta$ because the input interactions are also state dependant. The same reasoning applies to formulate the four inequalities arising from the binary product $\alpha_{pi}(1-\beta_{pk})$ in (\ref{eqn12}).

\begin{table}[ht]
\centering
\normalsize
\caption{Auxiliary variable $\delta$}
\begin{tabular}{ccccc}
\toprule[1.5pt]
\multicolumn{3}{c}{\textbf{Primary}} & {\textbf{Auxiliary}} \\
$\alpha_{pi}$ & $\beta_{pk}$ & $\alpha_{pi}(1-\beta_{pk})$ & $\delta_{pik}$ \\
\cmidrule[0.4pt](lr){1-3}\cmidrule[0.4pt](lr){4-4}
0 & 0 & 0 & 0 \\
0 & 1 & 0 & 0 \\
1 & 0 & 1 & 1 \\
1 & 1 & 0 & 0 \\
\bottomrule[1.5pt]
\end{tabular}
\\[5pt]
\end{table}

The four linear constraints associated to $\delta$ are represented in (\ref{eqn14}).

\begin{subequations} \label{eqn14}
$\forall (p;i;k) \in \llbracket 1; P \rrbracket \times \llbracket 1; N \rrbracket \times \llbracket 1; M \rrbracket,$
\begin{align}  \label{eqn14a}
\delta_{pik} \leq \alpha_{pi} + \beta_{pk} \\ \label{eqn14b}
\delta_{pik} \leq 1 + \alpha_{pi} - \beta_{pk} \\ \label{eqn14c}
\delta_{pik} \geq \alpha_{pi} - \beta_{pk} \\ \label{eqn14d}
\delta_{pik} \leq 2 - \alpha_{pi} - \beta_{pk} 
\end{align}
\end{subequations}

Both auxiliary variables have three indexes and can be represented by cubic arrays of binary variables, their respective sizes are $P \times N \times N$ for the state auxiliary variable $\gamma$ and $P \times N \times M$ for the input auxiliary variable $\delta$. In addition to the constraints presented in (\ref{eqn13}) and (\ref{eqn14}), both auxiliary variables have to be composed only of binary variables.

The optimization problem presented in (\ref{eqn12}) is reformulated into the linear integer optimization problem presented in (\ref{eqn15}), obtained by replacing the binary products by the auxiliary variables along with their constraints.

\begin{equation} \label{eqn15}
\begin{aligned}
&\underset{\alpha ,\beta}{\text{min}} & & \displaystyle \sum_{p=1}^{P} \sum_{i=1}^{N} \bigg\{ \sum_{j=1}^{N} \big\{ \gamma_{pij} \lvert a_{ij} \rvert \big\} + \sum_{k=1}^{M} \big\{ \delta_{pik} \lvert b_{ik} \rvert \big\} \bigg\} \\
&\text{s. t.}
& & (\ref{eqn5}),(\ref{eqn13}),(\ref{eqn14})
\end{aligned}
\end{equation}

As it was presented within this section, the use of two auxiliary variables enables the linearisation of the optimization problem. The minimization problem presented in (\ref{eqn15}) because of the use of primary and auxiliary binary variables allows to trade non-linearities for an increase in variable size. Nonetheless the complexity of the optimization problem can be reduced by exploiting the structure of the plant model and by creating only the auxiliary variables where the state space model elements are not equal to zero. For instance, if $a_{ij}=0$ there is no need to create $(\gamma_{pij})_{p \in \llbracket 1;P \rrbracket}$, similarly if $b_{ik}=0$ with $(\delta_{pik})_{p \in \llbracket 1;P \rrbracket}$. The next section presents the notion of controllability cuts reducing the search space in order to obtain only controllable subsystems.

\section{Controllability cut}

Running the previous optimization problem will yield $P$ subsystems presenting the least amount of interactions, unfortunately no information is given concerning their controllability. The state space model of any subsystem $p$, represented without the couplings coming from the other subsystems can be rewritten from (\ref{eqn2}) as follows

\begin{equation} \label{eqn16}
\forall p \in \llbracket 1; P \rrbracket, \, \dot{x}_{p} = A_{pp} x_{p} + B_{pp} u_{p}
\end{equation}

The controllability of any given subsystem model $p$ yielded by the optimization problem can be checked by verifying that the controllability matrix $C_{p}$ defined in (\ref{eqn17}) has full row rank.

\begin{equation} \label{eqn17}
\forall p \in \llbracket 1; P \rrbracket, \, C_{p} = [B_{pp} | A_{pp}B_{pp} | \ldots | A^{N_{p}-1}_{pp}B_{pp}]
\end{equation}

Therefore, at the end of the optimization process, a controllability test is performed for each subsystem model, testing that the set of equalities given in (\ref{eqn18}) holds.

\begin{equation} \label{eqn18}
\forall p \in \llbracket 1; P \rrbracket, \, rank(C_{p}) = N_{p}
\end{equation}

As one can see the controllability matrices $C_{p}$ as well as the integers $N_{p}$ representing the number of states in subsystem $p$ are results of the optimization and are not known a priori. Therefore implementing constraints within the linear integer optimization problem in order to restrain the solutions to the set of controllable subsystems is a tremendously difficult task. However, applying controllability cuts to the search space recursively and a posteriori is possible.

Every time a non-controllable partitioning is achieved new linear constraints are added to the existing ones in order to reduce the search space by cutting the non-controllable partitionings out with an affine hyperplane. The principle of cutting solutions out of the search space is similar to the Gomory cuts (\cite{Gomory1958}) where cuts are used to discard solutions that are not integer. Controllability cuts are applied from the root node and are valid for the entire search tree, hence cuts lifting methods are not necessary in this case (\cite{Balas1996}).

Grouping matrices can be seen as a concatenation of basis vectors, such that

\begin{subequations} \label{eqn4}
\begin{align} \label{eqn4a}
\alpha = [e_{i_{1}} | e_{i_{2}} | \ldots | e_{i_{k}} | \ldots | e_{i_{N}}]_{i_{k} \in \llbracket 1;P \rrbracket} \\ \label{eqn4b}
\beta = [e_{i_{1}} | e_{i_{2}} | \ldots | e_{i_{k}} | \ldots | e_{i_{M}}]_{i_{k} \in \llbracket 1;P \rrbracket}
\end{align}
\end{subequations}

with $(e_{i_{k}})_{i_{k} \in \llbracket 1;P \rrbracket}$ the canonical orthonormal basis of $\mathbb{R}^{P}$. Subsequently the square of their 2-norm is equal to respectively $N$ and $M$ as it is calculated in (\ref{eqn19}) (\ref{eqn20}).

\begin{equation} \label{eqn19}
\lVert \alpha \rVert^{2}_{2} = trace(\alpha^{\intercal} \alpha) = \displaystyle \sum_{k=1}^{N} e^{\intercal}_{i_{k}}.e_{i_{k}} = \displaystyle \sum_{k=1}^{N} \delta_{i_{k}i_{k}} = N
\end{equation}

\begin{equation} \label{eqn20}
\lVert \beta \rVert^{2}_{2} = trace(\beta^{\intercal} \beta) = \displaystyle \sum_{k=1}^{M} e^{\intercal}_{i_{k}}.e_{i_{k}} = \displaystyle \sum_{k=1}^{M} \delta_{i_{k}i_{k}} = M
\end{equation}

The set of non-overlapping grouping matrices of size $P \times N$ and respecting the constraints (\ref{eqn5a}) (\ref{eqn5c}) (\ref{eqn5e}) will be referred to as $\mathbb{G}_{PN}$ with $\mathbb{G}_{PN} \subset \llbracket 0; 1 \rrbracket^{P \times N}$. For a given non-controllable optimal partitioning denoted by $\alpha^{nc^{*}}$ and $\beta^{nc^{*}}$, and for any couple of non-overlapping grouping matrices $\alpha$ and $\beta$, the inequalities (\ref{eqn21}) and (\ref{eqn22}) hold.

\begin{align} \label{eqn21}
& \forall (\alpha^{nc^{*}},\alpha) \in \mathbb{G}_{PN}^{2}, \nonumber \\ 
& trace(\alpha^{nc^{*}\intercal} \alpha) = \displaystyle \sum_{k=1}^{N} e^{\intercal}_{i_{k^{nc^{*}}}}.e_{i_{k}} = \displaystyle \sum_{k=1}^{N} \delta_{i_{k^{nc^{*}}}i_{k}} \leq N
\end{align}

\begin{align} \label{eqn22}
& \forall (\beta^{nc^{*}},\beta) \in \mathbb{G}_{PM}^{2}, \nonumber \\ 
& trace(\beta^{nc^{*}\intercal} \beta) = \displaystyle \sum_{k=1}^{M} e^{\intercal}_{i_{k^{nc^{*}}}}.e_{i_{k}} = \displaystyle \sum_{k=1}^{M} \delta_{i_{k^{nc^{*}}}i_{k}} \leq M
\end{align}

In addition, the upper bound is only reached in (\ref{eqn21}) when $\alpha = \alpha^{nc^{*}}$ and in (\ref{eqn22}) when $\beta = \beta^{nc^{*}}$. Consequently there exists a natural way of constructing affine cutting hyperplanes (\ref{eqn23}) when a non-controllable optimal partitioning $(\alpha^{nc^{*}},\beta^{nc^{*}})$ is obtained.

\begin{subequations} \label{eqn23}
\begin{gather}
\forall (\alpha,\beta) \in \mathbb{G}_{PN} \times \mathbb{G}_{PM}, \, (\alpha,\beta) \neq (\alpha^{nc^{*}},\beta^{nc^{*}}) \nonumber \\ \label{eqn23a}
\Leftrightarrow trace(\alpha^{nc^{*}\intercal} \alpha) + trace(\beta^{nc^{*}\intercal} \beta) \leq N + M - 1 \\ \label{eqn23b}
\Leftrightarrow \displaystyle \sum_{p=1}^{P} \bigg\{ \sum_{i=1}^{N} \big\{ \alpha^{nc^{*}}_{pi} \alpha_{pi} \big\} + \sum_{k=1}^{M} \big\{\beta^{nc^{*}}_{pk} \beta_{pk} \big\} \bigg\} \nonumber \\
\leq N + M - 1
\end{gather}
\end{subequations}

Every time a non-controllable partitioning is obtained a controllability cut (\ref{eqn23b}) is added to the linear constraints before the optimization is computed again. Therefore, the previous non-controllable optimal partitioning can no longer be reached as it is now excluded from the search space. However because the groups are not ordered the same result can be achieved again simply by swapping the rows of $\alpha$ and $\beta$, leading to another representation of the same partitioning. Indeed, without any order constraints on the groups, $P!$ identical representations of a single partitioning are possible.

Different techniques can be employed to make the optimization more efficient. First of all, it would be possible to constraint the grouping matrices in order to rank the different groups as it has been done with move blocking matrices (\cite{Cagienard2007}). In this case only one representation per partitioning would be possible. The second solution would be to perform $P!$ controllability cuts every time a non-controllable optimal solution is obtained. Therefore all the possible non-controllable representations of a partitioning would be removed from the search space simultaneously. It is the latter solution that has been implemented in the weak interactions partitioning algorithm. Every time an optimal non-controllable solution is encountered $P!$ controllability cuts are added to the linear set of existing constraints. The optimization is ran iteratively until the least interacting controllable partitioning is found. The next section presents how the partitioning algorithm is built using the linear relaxation as well as the controllability cut technique.

\section{Partitioning algorithm}

The algorithm implemented to perform the system partitioning takes the main state space model as well as the group number $P$ as inputs. It returns the state and input grouping matrices $\alpha$ and $\beta$ once one of the least interacting controllable partitioning is reached. The algorithm can be described by steps as it is computed. The first step is to build the linear constraints that will be used for the primary and auxiliary variables. Then the optimization is performed yielding the subsystem models as well as their controllability matrices. The last step of the algorithm is to add the appropriate set of controllability cuts if the optimal partitioning presents at least one uncontrollable subsystem as well as to go back to the previous step. Otherwise, the algorithm returns the least interacting controllable partitioning as a final result if the optimal partitioning obtained has all its subsystem models controllable.

The weak interaction partitioning problem is a 0-1 integer linear programming problem. The partitioning algorithm is presented below in algorithm \ref{alg1}.

\setlength{\algoheightrule}{1.5pt}
\setlength{\algotitleheightrule}{0.4pt}
\SetAlgoNoLine
\begin{algorithm}[ht]
	\SetKwInOut{Input}{Input}
    \SetKwInOut{Output}{Output}

    \Input{$A,B,P$}
    \Output{$\alpha,\beta$}
    
    \While{$\exists p \in \llbracket 1; P \rrbracket, \, rank(C_{p}) \neq N_{p}$}{
    	
    	\eIf{$\exists (\alpha^{nc^{*}},\beta^{nc^{*}})$}{
    	Add controllability cuts (\ref{eqn23b})\\
    	Run the optimization problem (\ref{eqn15}) subject to (\ref{eqn23})\\
    	Extract the subsystem state space models and compute: $\forall p \in \llbracket 1; P \rrbracket, \, C_{p}$
    	}
    	{
    	Run the optimization problem (\ref{eqn15})\\
    	Extract the subsystem state space models and compute: $\forall p \in \llbracket 1; P \rrbracket, \, C_{p}$
    	}
    }
\caption{Partitioning algorithm}
\label{alg1}
\end{algorithm}

On the very first loop iteration, no solution exists, therefore, the optimization is performed and the first grouping matrices are obtained. The next step is to check that every subsystem is controllable by verifying that (\ref{eqn18}) holds. If at least one of the subsystems is not controllable then controllability cuts are added to the constraints and the optimization can start again using the reduced search space.
The algorithm finishes when the least interacting controllable partitioning comprising $P$ groups is found or when no controllable partitioning can be established.

\section{Examples}

This section presents some examples that were used to test the partitioning algorithm. The first example presented was used only to test the linear optimization and does not require any controllability cuts. However the second example was used specifically to demonstrate the controllability check and the controllability cuts.

\subsection{Example without controllability cuts}

\begin{figure*}[!t]
\begin{subequations} \label{eqn24}
\begin{align} \label{eqn24a}
A &= \left(
\begin{array}{ccccc}
-0.3245 \times 10^{1} & -0.2158 \times 10^{1} & -0.9155 \times 10^{3} & 0.5731 \times 10^{0} & 0.1342 \times 10^{3} \\ 
0.1642 \times 10^{1} & -0.5941 \times 10^{1} & -0.2816 \times 10^{3} & 0.1897 \times 10^{0} & 0.5705 \times 10^{2} \\ 
0.1685 \times 10^{-1} & -0.2554 \times 10^{-1} & -0.1003 \times 10^{2} & 0.7994 \times 10^{-2} & 0.5807 \times 10^{0} \\ 
0 & 0 & 0 & -0.1 \times 10^{2} & 0 \\ 
-0.2163 \times 10^{1} & 0.6862 \times 10^{1} & 0.7405 \times 10^{3} & 0.1195 \times 10^{1} & -0.1715 \times 10^{3}
\end{array}
\right)
\\ \label{eqn24b}
B &=\left(
\begin{array}{ccccc}
0.1432 \times 10^{-1} & -0.3553 \times 10^{3} & -0.9906 \times 10^{2} & -0.1549 \times 10^{2} & 0.222 \times 10^{5} \\
0.2871 \times 10^{0} & 0.7286 \times 10^{3} & 0.2514 \times 10^{2} & -0.6487 \times 10^{2} & 0.8122 \times 10^{4} \\
-0.2469 \times 10^{-2} & -0.103 \times 10^{3} & 0.6333 \times 10^{0} & -0.3213 \times 10^{0} & -0.7418 \times 10^{2} \\
0.1 \times 10^{2} & 0 & 0 & 0 & 0 \\
-0.1311 \times 10^{0} & 0.3295 \times 10^{3} & -0.25 \times 10^{2} & 0.6257 \times 10^{2} & -0.6445 \times 10^{5}
\end{array}
\right)
\\ \label{eqn24c}
P &= 2
\end{align}
\end{subequations}
\end{figure*}

The first example tested is the state space model of a military engine, the Pratt and Whitney $F100$ taken from (\cite{Jaw2009}). The algorithm was used with the parameters given in (\ref{eqn24}). The reader can see that the whole system is already controllable even before performing any kind of partitioning.

The partitioning obtained can be guessed due to the presence of zeros but also because of the presence of large elements in the matrices. The two grouping matrices resulting from the optimization are given in (\ref{eqn25}). It is important to notice that this first example does not need any controllability cut.

\begin{subequations} \label{eqn25}
\begin{align} \label{eqn25a}
\alpha &= \left(
\begin{array}{ccccc}
 0 & 0 & 0 & 1 & 0 \\
 1 & 1 & 1 & 0 & 1
\end{array}
\right)
\\ \label{eqn25b}
\beta &= \left(
\begin{array}{ccccc}
 1 & 0 & 0 & 0 & 0 \\
 0 & 1 & 1 & 1 & 1
\end{array}
\right)
\end{align}
\end{subequations}

This example has been run on a standard desktop with an execution time of $0.54s$.

\subsection{Example involving controllability cuts}
The second example is defined such that

\begin{subequations} \label{eqn26}
\begin{align}
A &= \left(
\begin{array}{ccccc}
1 & 1 & 0 & 0 & 0 \\ 
1 & -1 & 0 & 0 & 0 \\ 
0 & 0 & 1 & 1 & 0 \\ 
0 & 0 & 1 & 1 & 0 \\
0 & 0 & 0 & 0 & -1 \\
\end{array}
\right)
\\
B &= \left(
\begin{array}{ccccc}
1 & 0 & 0 & 1 & 0 \\ 
1 & 0 & 0 & 1 & 0 \\ 
0 & 1 & 0 & 0 & 1 \\ 
0 & 1 & 0 & 0 & 1 \\
0 & 0 & 1 & 0 & 0 \\
\end{array}
\right)
\\
P &= 3
\end{align}
\end{subequations}

The first $19$ iterations of the algorithm result in non-controllable partitionings. After performing $114$ controllability cuts, being $19 \times 3!$, the least interacting controllable partitioning is obtained (\ref{eqn27}).

\begin{subequations} \label{eqn27}
\begin{align}
\alpha &= \left(
\begin{array}{ccccc}
 0 & 0 & 0 & 1 & 0 \\
 1 & 1 & 1 & 0 & 0 \\
 0 & 0 & 0 & 0 & 1
\end{array}
\right)
\\
\beta &= \left(
\begin{array}{ccccc}
 0 & 0 & 0 & 0 & 1 \\
 1 & 1 & 0 & 1 & 0 \\
 0 & 0 & 1 & 0 & 0
\end{array}
\right)
\end{align}
\end{subequations}

For three groups each controllability cut has to be performed $6$ times in order to take into account all the possible permutations. On a standard desktop the total running time was $15.4s$.

\section{Conclusion}

An integer linear programming approached has been presented within this paper to tackle the problem of partitioning a system model into non-overlapping subsystem models. Auxiliary binary variables have been introduced in order to linearise the objective function. Finally, a method similar to Gomory cuts has been implemented in order to rule out the least interacting partitionnings including at least one non-controllable subsystem model. Because the partitioning problem is a combinatorial problem, the size of the search space increases very rapidly with the size of the system and the number of groups. Therefore the computational cost is important for large scale systems. Future work will address studying the performance of the algorithm as well as its complexity.

\bibliography{root}             

\begin{thebibliography}{24}
\providecommand{\natexlab}[1]{#1}
\providecommand{\url}[1]{\texttt{#1}}
\providecommand{\urlprefix}{URL }
\expandafter\ifx\csname urlstyle\endcsname\relax
  \providecommand{\doi}[1]{doi:\discretionary{}{}{}#1}\else
  \providecommand{\doi}{doi:\discretionary{}{}{}\begingroup
  \urlstyle{rm}\Url}\fi

\bibitem[{Bakule(2008)}]{Bakule2008}
Bakule, L. (2008).
\newblock Decentralized control: An overview.
\newblock \emph{Annual Reviews in Control}, 32, 87--98.

\bibitem[{Balas et~al.(1996)Balas, Ceria, Cornu\'{e}jols, and
  Natraj}]{Balas1996}
Balas, E., Ceria, S., Cornu\'{e}jols, G., and Natraj, N. (1996).
\newblock Gomory cuts revisited.
\newblock \emph{Operations Research Letters}, 19, 1--9.

\bibitem[{Bemporad and Morari(1999)}]{Bemporad1999}
Bemporad, A. and Morari, M. (1999).
\newblock Control of systems integrating logic, dynamics, and constraints.
\newblock \emph{Automatica}, 35, 407--427.

\bibitem[{Bristol(1966)}]{Bristol1966}
Bristol, E.H. (1966).
\newblock On a new measure of interaction for multivariable process control.
\newblock \emph{IEEE Transactions on Automatic Control}, 11, 133--134.

\bibitem[{Browning(2001)}]{Browning2001}
Browning, T.R. (2001).
\newblock Applying the design structure matrix to system decomposition and
  integration problems: A review and new directions.
\newblock \emph{IEEE Transactions on Engineering Management}, 48, 292--306.

\bibitem[{Cagienard et~al.(2007)Cagienard, Grieder, Kerrigan, and
  Morari}]{Cagienard2007}
Cagienard, R., Grieder, P., Kerrigan, E.C., and Morari, M. (2007).
\newblock Move blocking strategies in receding horizon control.
\newblock \emph{Journal of Process Control}, 17, 563--570.

\bibitem[{Cavalier et~al.(1990)Cavalier, Pardalos, and Soyster}]{Cavalier1990}
Cavalier, T.M., Pardalos, P.M., and Soyster, A.L. (1990).
\newblock Modeling and integer programming techniques applied to propositional
  calculus.
\newblock \emph{Computers \& Operations Research}, 17, 561--570.

\bibitem[{Chen and Seborg(2003)}]{Chen2003}
Chen, D. and Seborg, D.E. (2003).
\newblock Design of decentralized pi control systems based on {N}yquist
  stability analysis.
\newblock \emph{Journal of Process Control}, 13, 27--39.

\bibitem[{Gomory(1958)}]{Gomory1958}
Gomory, R.E. (1958).
\newblock Outline of an algorithm for integer solutions to linear programs.
\newblock \emph{Bulletin of the American Mathematical Society}, 64, 275--278.

\bibitem[{Jamoom et~al.(1998)Jamoom, Feron, and McConley}]{Jamoom1998}
Jamoom, M.B., Feron, E., and McConley, M.W. (1998).
\newblock Optimal distributed actuator control grouping scheme.
\newblock \emph{Proceedings of the $37^{th}$ IEEE Conference on Decision \&
  Control}, 2, 1900--1905.

\bibitem[{Jaw and Mattingly(2009)}]{Jaw2009}
Jaw, L.C. and Mattingly, J.D. (2009).
\newblock \emph{Aircraft Engine Controls: Design, System Analysis, and Health
  Monitoring}.
\newblock AIAA, Reston.

\bibitem[{Kariwala et~al.(2003)Kariwala, Forbes, and Meadows}]{Kariwala2003}
Kariwala, V., Forbes, J.F., and Meadows, E.S. (2003).
\newblock Block relative gain: properties and pairing rules.
\newblock \emph{Industrial \& Engineering Chemistry Research}, 42, 4564--4574.

\bibitem[{Leininger(1979)}]{Leininger1979}
Leininger, G.G. (1979).
\newblock Diagonal dominance for multivariable {N}yquist array methods using
  function minimisation.
\newblock \emph{Automatica}, 15, 339--345.

\bibitem[{Maciejowski(2002)}]{Maciejowski2002}
Maciejowski, J.M. (2002).
\newblock \emph{Predictive control with constraints}.
\newblock Prentice Hall, Harlow.

\bibitem[{Manousiouthakis et~al.(1986)Manousiouthakis, Savage, and
  Arkun}]{Manousiouthakis1986}
Manousiouthakis, V., Savage, R., and Arkun, Y. (1986).
\newblock Synthesis of decentralized process control structures using the
  concept of block relative gain.
\newblock \emph{American Institute of Chemical Engineers}, 32, 991--1003.

\bibitem[{Mayne(2014)}]{Mayne2014}
Mayne, D.Q. (2014).
\newblock Model predictive control: Recent developments and future promise.
\newblock \emph{Automatica}, 50, 2967--2986.

\bibitem[{Moro\c{s}an et~al.(2010)Moro\c{s}an, Bourdais, Dumur, and
  Buisson}]{Morosan2010}
Moro\c{s}an, P.D., Bourdais, R., Dumur, D., and Buisson, J. (2010).
\newblock Building temperature regulation using a distributed model predictive
  control.
\newblock \emph{Energy and Buildings}, 42, 1445--1452.

\bibitem[{Motee and Sayyar-Rodsari(2003)}]{Motee2003}
Motee, N. and Sayyar-Rodsari, B. (2003).
\newblock Optimal partitioning in distributed model predictive control.
\newblock \emph{Proceedings of the American Control Conference}, 6, 5300--5305.

\bibitem[{Rawlings and Mayne(2009)}]{Rawlings2009}
Rawlings, J.B. and Mayne, D.Q. (2009).
\newblock \emph{Model Predictive Control : Theory and Design}.
\newblock Nob Hill Publishing, Madison.

\bibitem[{Scattolini(2009)}]{Scattolini2009}
Scattolini, R. (2009).
\newblock Architectures for distributed and hierarchical model predictive
  control –- {A} review.
\newblock \emph{Journal of Process Control}, 19, 723--731.

\bibitem[{Schuler et~al.(2014)Schuler, M\"{u}nz, and
  Allg\"{o}wer}]{Schuler2014}
Schuler, S., M\"{u}nz, U., and Allg\"{o}wer, F. (2014).
\newblock Decentralized state feedback control for interconnected systems with
  application to power systems.
\newblock \emph{Journal of Process Control}, 24, 379--388.

\bibitem[{Stewart et~al.(2010)Stewart, Venkat, Rawlings, Wright, and
  Pannocchia}]{Stewart2010}
Stewart, B.T., Venkat, A.N., Rawlings, J.B., Wright, S.J., and Pannocchia, G.
  (2010).
\newblock Cooperative distributed model predictive control.
\newblock \emph{Systems \& Control Letters}, 59, 460--469.

\bibitem[{\v{S}iljak(1991)}]{Siljak1991}
\v{S}iljak, D.D. (1991).
\newblock \emph{Decentralized control of complex systems}.
\newblock Academic Press Inc., San Diego.

\bibitem[{Williams(2013)}]{Williams2013}
Williams, P.H. (2013).
\newblock \emph{Model building in mathematical programming}.
\newblock Wiley, Chichester.

\end{thebibliography}
%
%
%
%
%
%

\end{document}